# Refining StreamBED through Expert Interviews, Design Feedback, and a Low Fidelity Prototype

**Alina Striner**

Human-Computer interaction Lab

iSchool | University of Maryland

College Park, MD 20740, USA

algol001@umd.edu

**Jennifer Preece**

Human-Computer interaction Lab

iSchool | University of Maryland

College Park, MD 20740, USA

preece@umd.edu

## Abstract

StreamBED is an embodied VR training for citizen scientists to make qualitative stream assessments. Early findings [11] garnered positive feedback about training qualitative assessment using a virtual representation of different stream spaces, but presented field-specific challenges; novice biologists had trouble interpreting qualitative protocols, and needed substantive guidance to look for and interpret environment cues. In order to address these issues in the redesign, this work uses research through design (RTD) methods to consider feedback from expert stream biologists, firsthand stream monitoring experience, discussions with education and game designers, and feedback from a low fidelity prototype. The qualitative findings found that training should *facilitate personal narratives*, *maximize realism*, and should use *social dynamics* to scaffold learning.

## ACM Classification Keywords

Virtual worlds training simulations; Software Design techniques; User Centered Design;

## Introduction

Citizen science crowdsourcing allows volunteers to collaborate with researchers in scientific data collection and analysis [1]. Environmental researchers spend years learning the nuances of collecting data, however volunteers often have limited experience in methodology, limited time, and limited training. The focus of this research is on teaching volunteers to understand and interpret the EPA's Rapid Bioassessment Protocol (RBP) [13], a qualitative measure of stream conditions (shown in Figure 1).

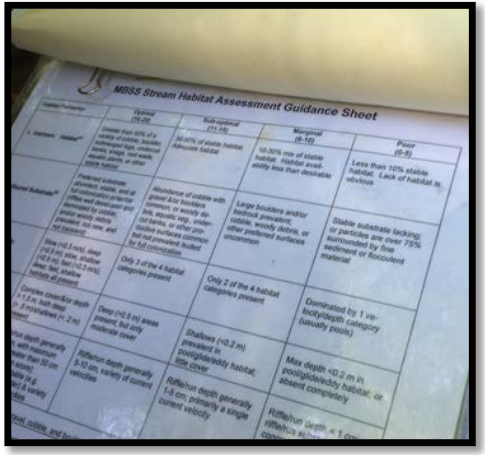

Figure 1: Physical version of the RPB protocol used during stream monitoring.

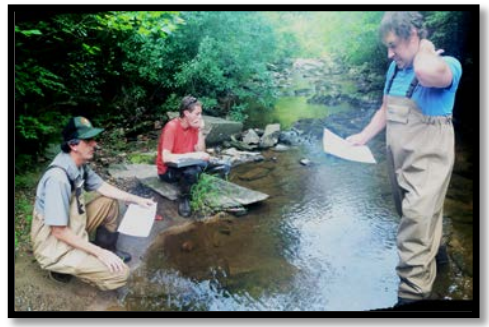

Figure 2: Group of field biologists evaluating a stream.

The goal of StreamBED is to design an optimal training for *qualitative* stream assessments. An alternative to traditional PowerPoint training [16], the platform takes advantage of current VR technologies, allowing non-expert audiences to experience on-site training at a fraction of time, cost, and human resources.

## Related Work
*Qualitative Judgments in Scientific Observations*
Biological observations consist of a process of noticing phenomena, comparing observations to expectations, and recording observations [4]. Noticing consists of knowing when to ask questions and what question to ask (e.g. "*what thing am I looking at?*"). After initially observing phenomena, scientists compare observations to established taxonomies or personalized webs of information [1]. In order to make deep observations, biologists create constructs from linked webs of information that can be chunked into personal models, and used to solve problems [15]. Holistic learning suits highly conceptual models with an underlying system, but does not help teach disconnected concepts or arbitrary rule-based systems; a web of interconnected ideas and metaphors are required for learners to examine, relate to, and consider questions fully [6].

*Pattern Recognition and Metaphors*
Intuition is a process of pattern recognition from past knowledge [3], observing and identifying similarities, differences and opportunities from a wealth of experience, and piecing together chunks of experience into a cognitive map [8]. Intuitive judgments are based on experiences that are subjective, difficult to surface, examine, and explain [3], and lead decision makers to interpret stimuli differently. Metaphors are a critical link between individual intuition and shared interpretation of meaning because they transfer information between familiar and new domains [3].

*Research Through Design*
In recent years, HCI has shifted from a narrow focus on usability to more broadly consider the human experience [1]. This expanded scope has lead to "wicked" problems, difficult research questions with unclear or conflicting agendas and messy solutions [1] [11] and limitless sources of information, requirements, and opportunities [1,18]. Research through design (RTD) methods support new knowledge creation through a design cycle of reflection and annotation [1], an iterative process of: *defining a problem, discovering and synthesizing data, generating, refining, and reflecting on solutions and evolving design*s [1]. The final output of a design is a concrete problem framing, and series of models, prototypes, and documentation of the design process [5]. This work uses the RTD method to comprehensively understand the methods of expert biologists and consider the needs of novice monitors.

## Designing StreamBED 2.0
The design goal of StreamBED 2.0 is to allow citizen scientists to learn qualitative monitoring in a way that naturally reflects the way expert monitors learn, but also supports non-expert needs for supplementary background, information scaffolding, and engagement mechanisms. An evaluation of the first prototype [11] found several challenges that needed to be addressed; Rather than reading the protocol, participants relied on misguided personal experiences to make stream judgments. Further, when prompted to follow the protocol, participants did not know how to interpret the RBP scale, needed concrete guidance to focus their attention on telling features of the environment, and expected more substantive feedback on their responses. The following section extends those initial findings by considering the needs of the redesign.

*Assessment Considerations*
The goal of StreamBED is to train non-experts to match expert observations and RBP assessments. In the field, RBP protocols are assessed together [14], however time constraints limit development to a subset of training protocols. Training must thus consider *which protocols* to focus on, and *which content cues* to use.

*Design Considerations*
Pilot participants had trouble focusing on relevant elements in the environment, and required more feedback and help. The redesign must thus consider the design of the *protocol cues, tasks and feedback*, *narrative and reward*, and *the user interface* needs described by Nielson's usability heuristics [1] and Preece's user experience goals [10]. Together, the interface should help participants smoothly and enjoyably navigate and interpret the virtual space.

*Attention, Feedback and Narrative.* It is also important to consider what attention and contextual cues and feedback should be included in the training design, and how they should be presented. Further, the design should have a *compelling narrative* that is entertaining, emotionally fulfilling, and motivates the training experience. Citizen scientists often come from varied backgrounds, so it is important that these elements support diverse backgrounds and needs.

## Expert and Design Feedback to Inform Design Framing

The focus of this work is to comprehensively consider the needs and practices of the water monitoring community in order to redesign StreamBED effectively. To achieve this goal, this work used RTD methods [1] to investigate how experts learn to do stream monitoring and conduct assessments. This process consisted of a series of expert interviews, on-site training and design feedback sessions that informed a low fidelity prototype design. Open coding [11] was used to identify meaningful trends in user responses; salient quotes were transcribed, grouped into overarching themes, and then organized into sub-themes. Themes identified in this research will inform the training redesign process.

*Expert Interviews and On-Site Stream Training*
In order to inform the study, the research interviewed 5 expert biologists either by phone or in person. In addition, the first author visited 6 stream sites with expert biologists, and received on site stream monitoring training (Figure 3). During this time, the first author recorded additional informal observations.

*Themes*
*Evaluating with the RBP*. The RBP protocol is a standard by which biologists do assessment, however they also develop an intuition for assessment quality. For instance, one senior biologist commented that "*after you've done this for 24 years...you can just sort of walk up to a stream, and at a glance, pretty much put it in a total category score, without even scoring it*." Interestingly, both senior biologists interviewed commented on having memorized variations on the protocol scales as sets of "*mental images*" from their years of experience. In contrast, an experience with a group of younger biologists suggested otherwise – they were not as comfortable with the protocol, and read the language aloud while making assessments.

*Additional Information.* Biologists were careful to stick to the language of the RBP while making assessments, however they used additional information to form opinions of a stream's quality. Several biologists mentioned supplementing the RBP with additional measures for trash, presence of human activity, and invasive plant and animal species. As an example, a biologist commented that hearing European Starlings (an invasive species) was indicative of poor stream quality. Likewise, another noted that invasive plant species may be "*a guide [that] ...the stream's probably not that intact*." Noticing these characteristics helped biologists look for stressors in the environment. Additionally, several biologists discussed using sensory information in their observations. For instance, one biologist approximated the strength of stream riffles by listening to the sound of rushing water, and another used sun warmth and wind strength to judge the quality of stream bank vegetation.

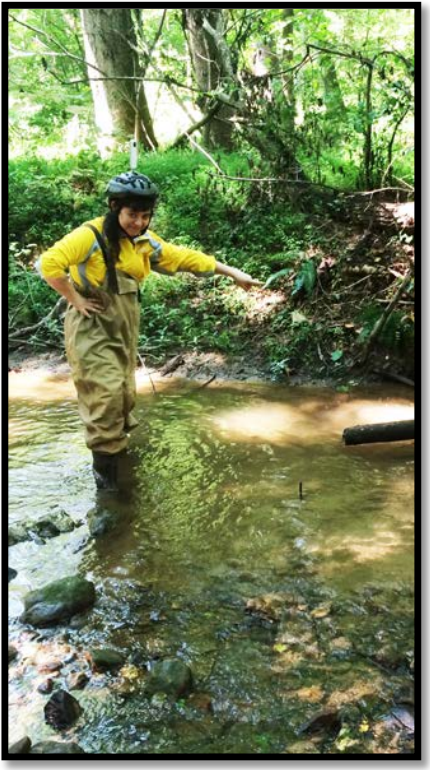

Figure 3: The first author making stream monitoring observations and capturing 360° images.

*Stream Narratives.* Through the RBP and observations, biologists actively interpreted the 'narrative' of a stream space: how it has changed, and what its likely to look like in the future. For instance, at a monitoring site in Virginia, a biologist pointed out that a stream used to be an old agriculture site, described the cause of a large erosion area, and predicted how the stream would change in the next 5-10 years. Similarly, describing the effect of sedimentation, one biologist emphasized how connected the ecosystem was, explaining that "*...simple things like keeping snakes out…they'll mow right out to the waters edge…the next time flow events... erode your whole bank.* " Rather than merely evaluating the present landscape, water biologists actively interpreted its change over time.

*Design Feedback Sessions*
Based on the expert interviews, the authors presented redesign suggestions and questions to three groups for feedback: two groups of HCI faculty and students, and one water monitoring group at the University of Maryland. The goal of these sessions was to garner suggestions from different parts of the HCI and water biology communities to inform the redesign.

*Themes*
*Story Narratives.* The design feedback sessions brought much debate on the role of narrative. When presented with 4 potential narratives, one group of HCI students argued whether narrative should be a "*thin framework*," "*different flavors of a scenario*," or whether it should be " *an immersive…story.*" Similarly, discordance arose over a dead fish narrative described in the low fidelity protocol below. One designer was excited by this story, explaining that "*you see like...a bunch of dead things at the end of the river, what are you gonna do?…[you] gotta go upriver...dead stuff is motivating.*" In contrast, a water monitoring faculty member worried about "*conflating a single event like fish kill,*" and explained that fish kills were "*usually just a characteristic of ...low dissolved oxygen, which is a result of …algal bloom.*"

*Realism.* Groups also focused heavily on the role of realism in scaffolded training and interactivity. Here, the monitoring faculty pointed out that "*you want them to take a look and then get a heuristic judgment of a particular area, and be able to consistently rank it… [regardless] of the event.*" The idea of realism also arose when discussing the learning process. For instance, a designer suggested using an avatar that reflected the role of an expert teacher, and another added that interactivity should use realistic metaphors. "*You want to give them tools that they're actually going to be able to use in the field...maybe if they pull out a compass, and the compass always pointed them in the direction of where they needed to go,*" they suggested.

**Low Fidelity Prototype Feedback**
The low fidelity prototype was a preliminary attempt to integrate feedback from expert water biologists and designers into a comprehensive design. Due to the issues described by biologists [11], this prototype replaced the visual protocol with image areas to focus on, environment descriptions, and relevant definitions.

During the study, 6 participants listened to a narrative of a journey to find the cause of dead fish downstream of a river. A PowerPoint slideshow supplied visual and audio cues about the state of each area of the stream, and participants received olfactory cues through scents sprayed into the air. A sample slide is shown in Figure 4 below.

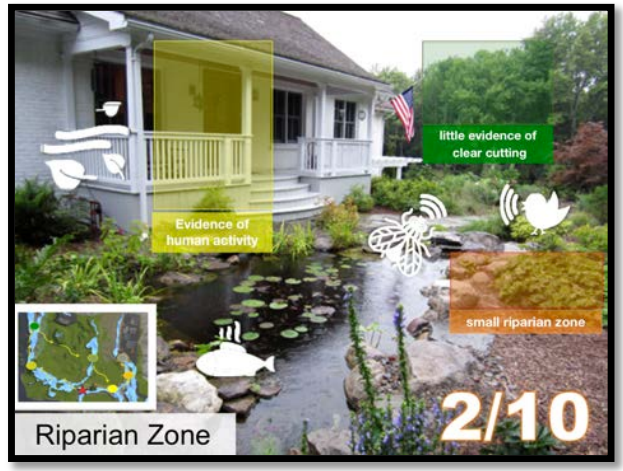

Figure 4: A sample stream image from the low fidelity study. The image highlights environment features related to the *Riparian Zone* protocol metric. The image includes an expert's scoring of the RBP metric and contextual cues: the location of the space on a minimap, and auditory and olfactory cues of birds, insects, and fish.

After the narrative training task, participants were asked to evaluate four images of stream environments using the protocols they had learned, and discussed their experiences as a group.

*Group Feedback*
*Detail Ambiguity.* The low fidelity training meant to present a low fidelity parallel to a VR prototype, however the lack of interactivity fundamentally limited its effectiveness. Participants explained that the visual presentation was not detailed enough, and that they needed step-by-step guidance to make sense of stream quality. For instance, one participant wasn't sure how "*how detailed...the training was*" and another clarified that " *[I understood]...certain keywords like 'covering' or shade as the presentation went on...but… it was hard to ...distinguish...all the different points*." Another specified that the ambiguity came from the areas highlighted in the images: *"...I couldn't quite figure out the [highlighted areas]…what part…was being highlighted or shown, cause I think …the [area was] too large."* To overcome this ambiguity, participants suggested quality markers that *"showed the actual erosion…a line, or… the depth, something that's going to make it easy to compare one picture to the next."*

*Inconsistency.* Additional problems came from inconsistency and incongruence of the environment and story. For instance, one person was dismayed by an image of a golf course stream." *I was thinking... you're camping along the Shenandoah… and …there are no golf courses in West Virginia*," they said. Another didn't feel that the narrative did not make sense: "*Why are the dead fish… [located downstream of]… a place that you character as being relatively okay?*" Although many issues were caused by secondary images and a lack of interactivity, the study made it clear that participants needed clear contextual information and guidance.

## Discussion

Through several in-depth-interviews, stream training sessions and a low fidelity study, this research found valuable design takeaways to incorporate into training.

*Personal narratives might be more effective than a forced storyline.* Several biologists, designers, and low fidelity participants noted that the proposed storylines were contrived, or didn't readily fit the stream assessment task. Rather than superimposing a narrative onto the user, feedback suggested that an effective narrative must be one that is constructed by the user. Not only did this work find that biologists naturally do this while doing stream assessment, but the pilot study found that participants generated stories that helped them explain virtual phenomena [11].

*The redesign should maximize high-fidelity resolution.* In-depth interviews found that biologists used multiple senses to make sense of stream spaces, and designers and low fidelity participants commented on the lack of detail in both the pilot prototype and PowerPoint images. Since RBP assessments require users to interact in multiple ways with the environment, training should maximize realism as much as possible. To do this, the redesigned training will use 360° images of real streams (Figure 5) that were captured on a Ricoh Theta camera.

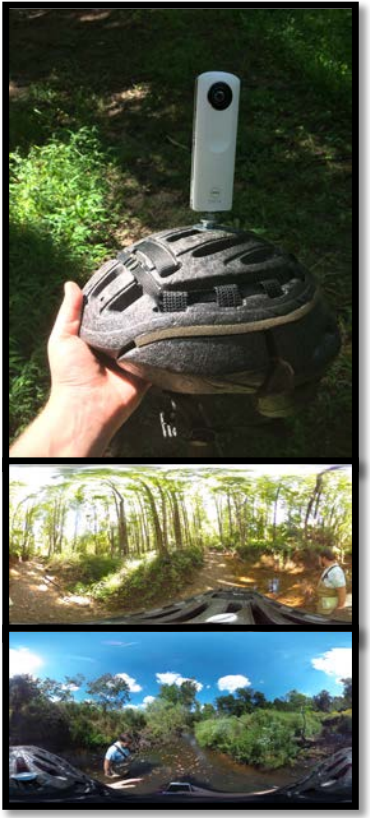

Figure 5: Ricoh Theta + helmet setup that was used to capture several 360° stream images, followed by sample images captured at stream sites.

*Negotiating protocol meaning should be as a social task.* During the interviews and stream experience, it was clear that interpreting the protocol was a social process of negotiation of meaning. All the stream surveys were conducted in groups (see Figure 2), and as one biologist noted, " *we always have pairs of 2…the person collecting the bugs is…able to answer [certain questions] because they're physically in the [stream]."* Similarly, designers suggested a social learning activity where a person in the VR environment could talk to a partner outside of VR; the person with the manual could guide their partner's exploration. This idea of social interpretation supports Crossan's work [3] on metaphors as negotiation of meaning; expressing intuition verbally helps users develop explicit connections between ideas, and begets further insights.

*Integration of Insights*

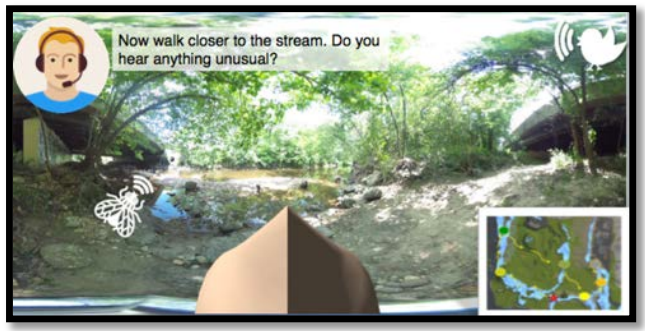

Figure 6: Mockup of a redesign with a natural storyline, high fidelity resolution with 360° images and multisensory information, and social learning with an avatar guide.

## Conclusions and Future Work

This research used a series of expert interviews, on-site training, design feedback sessions, and a low fidelity prototype to inform StreamBED 2.0, a high-fidelity redesign of water monitoring training for citizen scientists. Findings from the expert interviews, water monitoring experience, and feedback on the original prototype will inform the prototype redesign.

After iterating on the design of StreamBED 2.0, future research will compare the high fidelity prototype to a baseline PowerPoint experience [16] currently used to train citizen science water monitors. Future studies will evaluate whether embodied virtual training helps citizen scientists make qualitative stream judgments that approximate expert assessments and leads to higher engagement and motivation.

## Acknowledgements

Thanks to Dr. Gregory Pond, Dr. Jonathan Witt, Jeff Bailey, Carissa Turley, and Dr. Evan Golub for their generous time and feedback.